\documentclass[preprint,showpacs,preprintnumbers,amsmath,amssymb]{revtex4}

\usepackage{graphicx}
\usepackage{dcolumn}
\usepackage{bm}

\begin{document}

\title{Evolutionary dynamics of imatinib-treated leukemic cells by stochastic approach}
\author{\large Nicola Pizzolato$^{1*}$,
        Davide Valenti$^{2*}$,
        Dominique Persano Adorno$^3$,
        Bernardo Spagnolo$^{4*}$}
\affiliation{Dipartimento di Fisica e Tecnologie Relative, Universit\`a di Palermo and CNISM-INFM,\\
     Viale delle Scienze, edificio 18, I-90128 Palermo, Italy\\
     $^1$e-mail: npizzolato@gip.dft.unipa.it \enspace $^2$e-mail: valentid@gip.dft.unipa.it \enspace $^3$e-mail:
     dpersano@unipa.it \enspace $^4$e-mail: spagnolo@unipa.it\\$^*$URL: http://gip.dft.unipa.it}


\begin{abstract}The evolutionary dynamics of a system of cancerous cells
in a model of chronic myeloid leukemia (CML) is investigated by a
statistical approach. Cancer progression is explored by applying a
Monte Carlo method to simulate the stochastic behavior of cell
reproduction and death in a population of blood cells which can
experience genetic mutations. In CML front line therapy is
represented by the tyrosine kinase inhibitor imatinib which strongly
affects the reproduction of leukemic cells only. In this work, we
analyze the effects of a targeted therapy on the evolutionary
dynamics of normal, first-mutant and cancerous cell populations.
Several scenarios of the evolutionary dynamics of imatinib-treated
leukemic cells are described as a consequence of the efficacy of the
different modeled therapies. We show how the patient response to the
therapy changes when an high value of the mutation rate from healthy
to cancerous cells is present. Our results are in agreement with
clinical observations. Unfortunately, development of resistance to
imatinib is observed in a proportion of patients, whose blood cells
are characterized by an increasing number of genetic alterations. We
find that the occurrence of resistance to the therapy can be related
to a progressive increase of deleterious mutations.
\end{abstract}

\pacs{87.10Mn, 87.10.Rt, 87.23Kg, 87.19.xj\\Keywords: Stochastic
dynamics, Cancer evolution, Complex systems}

\maketitle


\section{Introduction}\label{intro}
\begin{figure}
\includegraphics[width=13.0cm]{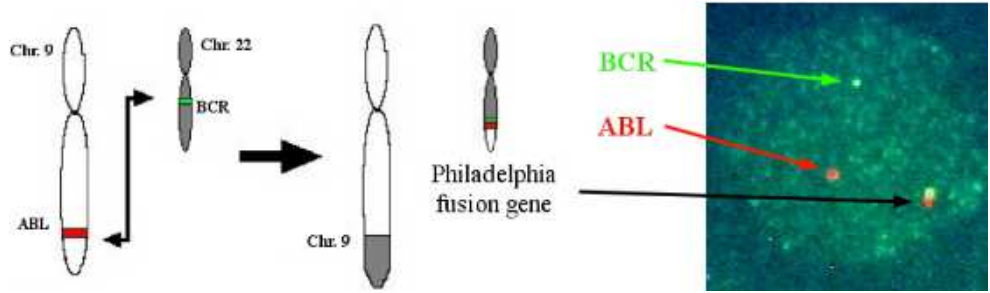}
\caption{Schematic representation of the Philadelphia translocation
(left) and detection mechanism by  Fluorescence Interphase in Situ
Hybridization (FISH) (right). \label{fig1}}
\end{figure}

In the last decade cancer dynamics and tumor growth models have been
attracted an increasing interest \cite{Michor2003,Michor04,Ywa2004,%
Michor2005,Abbo2006,Brum2006,Din2006,Fiasconaro06,Gar2006,Khodolenko06,%
Komar2006,Michor2006a,Michor2006b,Michor2006c,Roe2006,Komar2007,Chang08,%
Fiasconaro08,Zhdanov2008}. Among many different kind of human
cancer, a recent investigation of the dynamics of Chronic Myeloid
Leukemia (CML) has provided the first quantitative insights into the
\emph{in vivo} kinetics \cite{Michor2005}. In the bone marrow of
patients affected by CML too many myeloid cells (one of the main
types of white blood cells) are produced and released into the blood
when they are immature and unable to work properly. These immature
cells (called blastes) cannot do the work of normal white blood
cells, and this leads to an increased risk of infections.
Furthermore, they fill up the bone marrow making difficult the
production of enough healthy red cells and platelets (see
Ref.~\cite{Sawyers1999} for a recent review). CML cells are
characterized by a specific chromosomal abnormality: the
Philadelphia (Ph) chromosome \cite{Nowell1960, Rowley1973,
Shtivemman1985}, which is created by a reciprocal translocation
between part of the BCR ("breakpoint cluster region") gene from
chromosome 22 and part of the ABL gene on chromosome 9 (ABL stands
for "Abelson", the name of a leukemia virus which carries a similar
protein) (Fig. \ref{fig1}). BCR-ABL oncogene activates a number of
cell cycle-controlling proteins (p210 or sometimes p185) and enzymes
(tyrosine kinase), speeding up cell division. Moreover, it inhibits
DNA repairing, causing genomic instability.
Until recently, the only successful treatment of CML was to destroy
the patient's bone marrow and then restore blood-cell production by
infusing stem cells from the bone marrow of a healthy donor
\cite{McGlave1992, Clift1994, Savage1997}. In the late 1990s,
Novartis Pharmaceutical discovered a very efficient tyrosine kinase
inhibitor: STI-571 (imatinib, Gleevec/Glivec). Subsequent clinical
trials demonstrated that Gleevec inhibits the proliferation of
BCR-ABL-expressing hematopoietic cells, turning off the signal
produced by the Ph chromosome \cite{Druker1996, Hehlmann2000,
Talpaz2002, Cortes2005}. CML represents the first human cancer in
which molecularly targeted therapy leads to a striking clinical
response. Although it does not eradicate CML cells, it greatly
limits the growth of the tumor clone. Unfortunately, acquired
resistance to Gleevec develops in a substantial fraction of patients
\cite{Hochhaus2006}. The basis for resistance is a genetic change in
the BCR-ABL gene itself: in particular point mutations within the
protein tyrosine kinase domain \cite{Weis2001, Weis2003}.

The evolutionary dynamics of cancer beginning and progression has
been theoretically approached in several works with mathematical
deterministic equations \cite{Michor2005,Michor2006a,Abbo2006} or
with stochastic models
\cite{Michor2003,Ywa2004,Gar2006,Brum2006,Din2006,Komar2006,Michor2006b,%
Michor2006c,Komar2007,Zhdanov2008}. These works describe the
temporal evolution of the level of BCR-ABL positive cells,
experimentally observed in patients treated with Gleevec, in terms
of a partial or total failure of the drug efficacy on cancer stem
cells. This would explain the difficulty to completely eradicate the
cancer. Instead of a general insensitivity of stem cells to
imatinib, a selective effect on proliferative cells, as a process of
competition between healthy and cancerous stem cells for appropriate
niches, is also proposed \cite{Roe2006,Zhdanov2008}. On the other
hand, Gleevec effect on leukemic progenitors \cite{Michor2007}
appears to be strictly related to the rapid enhancement of the
number of blastes after a therapy stop.

In this work we investigate the evolutionary dynamics of leukemic
cells by simulating the stochastic behavior of cell division and
mutation in a system of initially normal blood cells which can be
affected by a double genetic mutation and transformed in cancerous
cells. In order to simulate the random process of cell selection for
reproduction, mutation and death (each process being guided by an
appropriate fitness or event rate) we adopt a Monte Carlo approach,
as already done by several authors in theoretical cancer studies
(see refs.
\cite{Berman1992,Delsanto2000,Ywa2004,Fiasconaro06,Michor2006a,%
Michor2006b,Michor2006c,Klein2007,Zhdanov2008}, to cite a few). The
aim of this work is to explain the several observed scenarios of CML
patient response to imatinib by modeling a therapy side effect of
enhancement in the number of deleterious mutations from healthy to
leukemic cells. In this framework, we also investigate the
development of resistance to the therapy and the subsequent relapse.
In sect.~\ref{sec1} we describe the model and give the details of
the simulation process; results are reported in sect.~\ref{res} and
conclusions are drawn in sect.~\ref{concl}.

\section{Stochastic model}\label{sec1}
The Ph-translocation is a particular form of chromosomal instability
(CIN) \cite{Leng1998,Clau2005}, that refers to increased rates of
gaining or losing whole chromosomes or arms of chromosomes. Several
hundred genes contribute to maintaining the stability of chromosomes
during cell division; mutations in such genes can trigger the CIN
phenotype. The basic idea that cancer arises when a single cell
experiences multiple mutations has been confirmed by numerous
studies on cancer genetics \cite{Moo1981,Knu2001,Fra2003}.

In this work we apply a Monte Carlo approach to study the dynamics
of a finite population of N replicating cells under the effect of
two sequential mutations. The total number of cells is kept constant
during the time evolution, as it can be reasonably assumed for a
blood cancer. In our simulations, we have chosen N$=10^4$ cells.
This value is several orders of magnitude lower than the typical
total contents of blood cells in humans, but it is great enough for
the statistical study of the cancer development in a single blood
compartment. Our model contains three types of cells, denoted by 0,
1, and 2. Cell population of type 0 can mutate to cells of type 1 at
a rate M$_{01}$ and type 1 cells mutate to type 2 at a rate
M$_{12}$. The reproductive rates (fitness) of cell types 0, 1 and 2
are labeled by F$_0$, F$_1$, F$_2$, respectively. In Fig. \ref{fig3}
is shown a logical map of the evolution of the blood cells from
normal to cancerous types. We do not consider back mutations and
neglect direct transitions from healthy (type 0) to leukemic cells
(type 2).

\begin{figure}
\includegraphics[width=10.0cm]{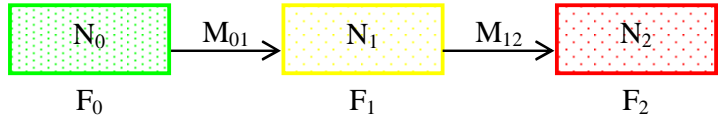}
\caption{Logical map of the evolution of the blood cells from normal
type (green box) to first-mutants (yellow box) and finally to
leukemic cells (red box).\label{fig2}}
\end{figure}

The stochastic dynamics of the cancer evolution is modeled by
assuming that cells reproduce asynchronously. This means that each
elementary step of the stochastic process consists of a birth and a
death event (Moran process \cite{Moran1962}). For the reproduction
process, one of the N cells is randomly chosen proportionally to the
fitness. It will give rise to an offspring (a new cell of the same
type) subject to possible mutation. The exact sequence of logical
steps for a single reproduction event is described by the flowchart
shown in Fig.~\ref{fig3}. Within the same time step a death event
occurs, being one of the N cells chosen at random to be eliminated
(see Fig.~\ref{fig4}). This guaranties that the total population
size remains constant. At time t=0, all N cells are of 0-type. After
some time, a single mutant cell of type 1 is generated. This cell
leads to a lineage of type 1 cells that could mutate to type 2 or go
to extinction before a second mutation event from type 1 to type 2
takes place. In order to give a statistical significance to our
descriptions, every simulation is repeated 500 times and all the
results reported in the next section are based on ensemble averages.
Time is measured in units of cell divisions. A time scaling,
described in the next section, is applied in order to show the
results in day time units.

\begin{figure}
\includegraphics[width=10.0cm]{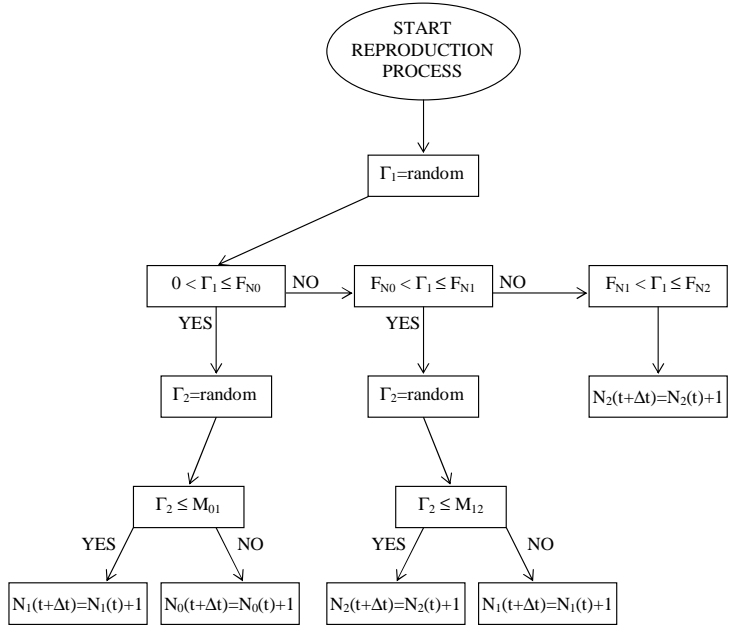}
\caption{Flowchart of the reproduction event in a single time step.
F$_{\rm{N}\it{i}}$ (with $i=0,1,2$) are the normalized fitness
rates; M$_{01}$, M$_{12}$ are mutation rates for transitions from
type 0 to type 1 cells and from type 1 to type 2 cells,
respectively. \label{fig3}}
\end{figure}

\begin{figure}
\includegraphics[width=8.0cm]{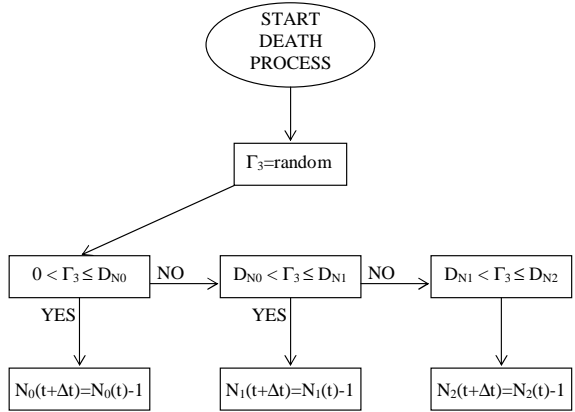}
\caption{Flowchart of the death event in a single time step.
D$_{\rm{N}\it{i}}$ (with $i=0,1,2$) are the normalized death rates.
\label{fig4}}
\end{figure}

\section{Results}\label{res}
\subsection{Therapy effects on cancer evolutionary dynamics}\label{ther}
\begin{figure}
\includegraphics[width=14.0cm]{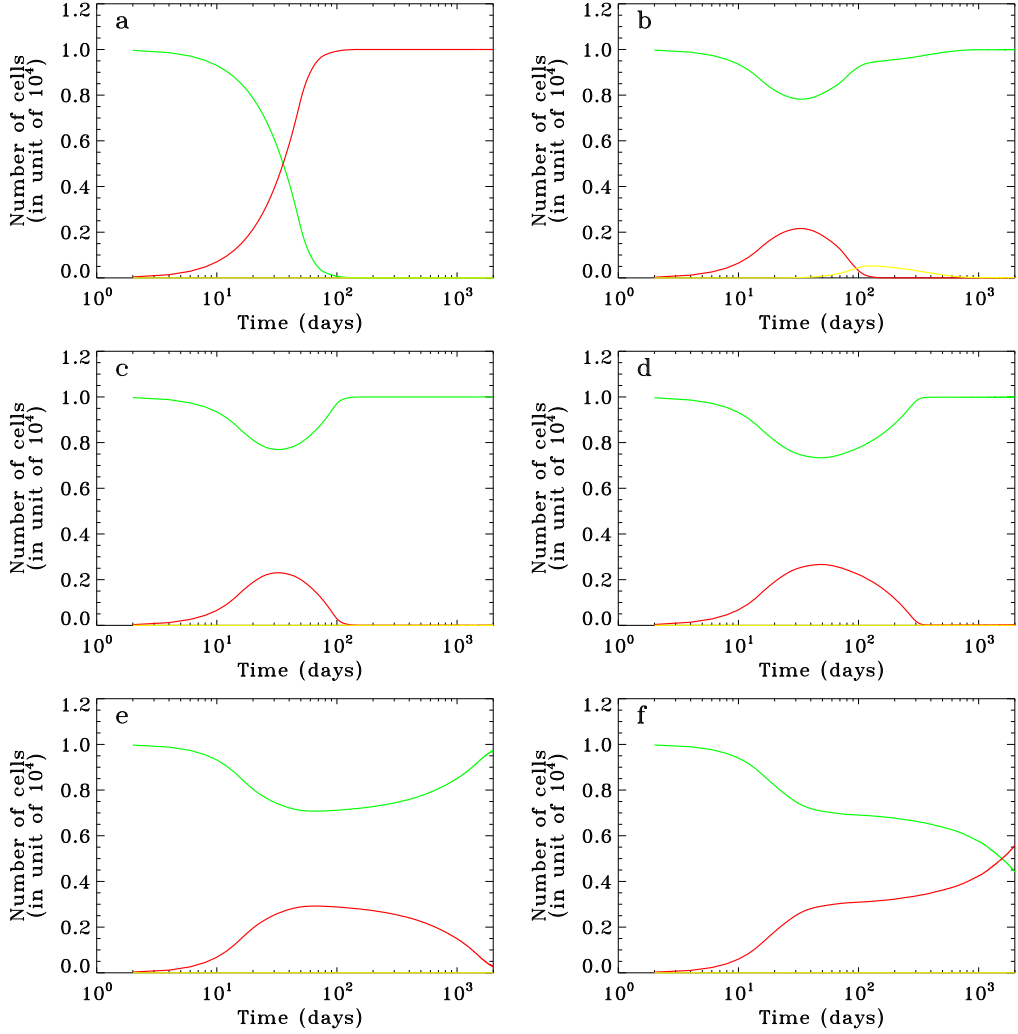}
\caption{Evolutionary dynamics of cancerous cells. The green line
indicates the behavior of healthy cells, yellow line that of first
mutation cells and red line that of second mutation cancerous cells:
(a) without any therapy; (b) including the effects of an ideal
therapy with an activation threshold of N$/3$ cancerous cells; (c,
d, e, f) a real therapy with increasing mutation rates (see table
\ref{tab1}).\label{fig5}}
\end{figure}

The first case under investigation is the stochastic evolution of
the three types of cell populations in the absence of any therapy.
Real values of mutation rates from normal cell to leukemic type are
not available because nobody knows with certainty how long is the
latency of the CML. We have chosen the mutation rates M$_{01}$ and
M$_{12}$ equal to 0.001 and 0.02, respectively. These values are one
order of magnitude greater than the mutation rates adopted elsewhere
\cite{Ywa2004} because, in this study, we are not interested in the
average waiting time before the disease make its first appearance,
but in the subsequent dynamic evolution. Normal cells (green line in
Fig.~\ref{fig5}a) experience the first mutation after some time
interval. If the intermediate-type cells (yellow line) survive for a
sufficient long time, a second-type mutation can cause the birth of
cancerous cells (red line). In our code the fitness F$_0$ and F$_1$
are set equal to 1, while the fitness of type 2 cells has been
reasonably assumed 10 times that of the other two populations. For
this reason, the number of cancerous cells rapidly increases to the
total initial value N of normal cells.

The effect of an ideal therapy is modeled by reducing the
reproductive capability of the only type 2 mutated cells. We know
that an imatinib-based therapy can also induce the CML cells
apoptosis \cite{Desplat2005,Belloc2007}, but, at this stage, we do
not include this effect in our model. Therefore the death rates
D$_0$, D$_1$ and D$_2$, associated to type 0, type 1 and type 2
cells, respectively, are left equal to each other. Since a
successful therapy requires a basic reproductive ratio of cancerous
cells lower than 1 \cite{Michor2006a}, we model the therapy effect
by lowering the fitness parameters F$_2$ to 0.7. All our simulations
start with the patient developing CML in the absence of therapy;
when the number of leukemic cells exceeds the threshold value N$/3$,
the therapy is activated. In fig.~\ref{fig5}b we can see that, after
an initial increase of the number of leukemic cells, the effect of
an ideal therapy is to completely eradicate the number of mutated
cells and favor the restoring of normal cells. The time scaling from
cell division to days is performed by assuming a complete restore of
healthy cells in almost 100 days, as experimentally observed in
clinical cases of optimal therapy response \cite{Michor2005,
Roe2006}.

For the study of real cases of therapy effects on cancer evolution,
we have supposed that the cell system could react to the drug
administration by activating multiple genetic changes that cause an
enhancement of the mutation rates from normal to cancerous cells.
This assumption is supported by the experimental evidence that
certain mutations increase the rate at which subsequent mutations
occur \cite{Jack1998, Loeb2002}. This secondary effect of the
therapy is investigated by modeling an increase of M$_{01}$  and
M$_{12}$ at four different levels, as summarized in table
\ref{tab1}. In the case 1 of real therapy (Fig.~\ref{fig5}c) we have
increased both mutation rates by a factor 10. While the extinction
effect on type 2 cells is almost unchanged with respect to the
previous case (Fig.~\ref{fig5}b), the first-mutant cells disappear
very quickly because of the increased value of M$_{12}$. When the
mutation rate M$_{12}$ is doubled and M$_{01}$  progressively
increased (cases 2, 3 and 4 in table \ref{tab1}), an effect of
retard is observed in the recovery of healthy cells
(Figg.~\ref{fig5}d and \ref{fig5}e) and, in the worst case, a
failure of the therapy itself is present (Fig.~\ref{fig5}f). This
means that, even if the therapy works properly by inhibiting the
reproduction of leukemic cells, the cancer proliferation can still
occur because of an increase of disadvantageous genetic mutations.

\begin{table}
\caption{Table of model parameters.\label{tab1}}
\begin{tabular}{lcccccc}
\hline\noalign{\smallskip}
& No therapy & Ideal therapy & Real therapy & Real therapy & Real therapy & Real therapy   \\
& & & (Case 1) & (Case 2) & (Case 3) & (Case 4) \\
\noalign{\smallskip}\hline\hline\noalign{\smallskip}
F$_2$ (cancerous cell) & 10.0 & 0.7 & 0.7 & 0.7 & 0.7 & 0.7 \\
M$_{01}$ & 0.001 & 0.001 & 0.01 & 0.10 & 0.13 & 0.14 \\
M$_{12}$ & 0.02 & 0.02 & 0.20 & 0.40 & 0.40 & 0.40 \\
\noalign{\smallskip}\hline
\end{tabular}
\end{table}

\subsection{Development of resistance to the therapy}\label{resist}
\begin{figure}[h]
\includegraphics[width=15.0cm]{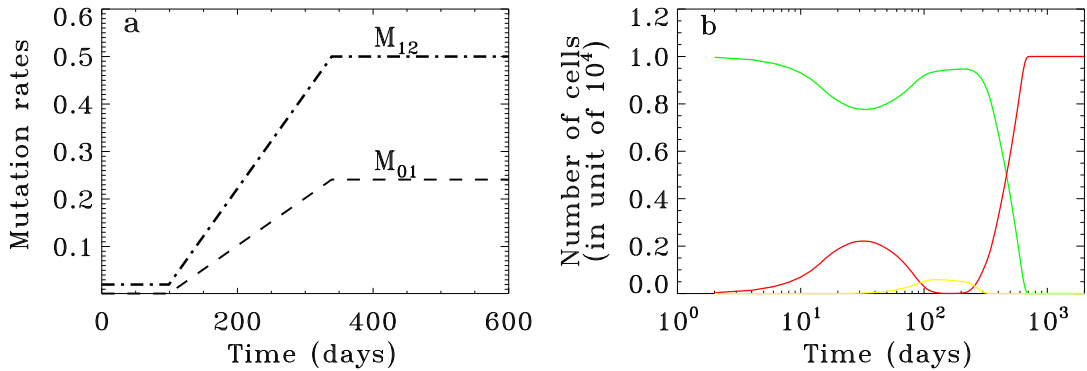}
\caption{(a) Changes on mutation rates with the time. (b)
Evolutionary dynamics of cancerous cells: development of resistance.
The green line indicates the behavior of healthy cells, yellow line
that of first mutation cells and red line that of second mutation
cancerous cells \label{fig6}}
\end{figure}

Several cases of acquired resistance to the therapy has been
observed in CML patients. In recent experimental works genetic
mutations are observed in a significant fraction of resistant
patients \cite{Sov2006}. An increase of the mutation rates could
represent the natural evolution of cancerous cells as a defense from
the action of the therapy itself. In the previous section we have
shown that a therapy based on the drug inhibitory action of leukemic
cell reproduction mechanism can fail, if the therapy also causes an
increase of the mutation rates from normal cells. We investigate the
development of resistance to the therapy by modeling a
time-dependent increase of both mutation rates. We have reasonable
decided to apply a time delay of 100 days, from the beginning of the
therapy, before M$_{01}$ and M$_{12}$ start to grow up. This value
has been chosen on the basis of experimental findings regarding an
initial response of the patient to the therapy before the first
appearance of the resistant clone. We have modeled a linear increase
of M$_{01}$ and M$_{12}$ with slopes of $0.001/$day and $0.002/$day,
respectively, for the subsequent 240 days, until M$_{01}$ reaches
the value 0.22 and M$_{12}$ 0.50. These values are assumed as upper
limits for M$_{01}$ and M$_{12}$, because greater mutation rates do
not significantly change the already dramatic cutoff of healthy
cells and the rapid increase of the cancerous cells. The time
dependent behavior of M$_{01}$ and M$_{12}$ are plotted in
Fig.~\ref{fig6}a. The evolutionary dynamics of the three cell
populations in CML patients, treated with a therapy causing a
progressive increase of mutation rates, are shown in
Fig.~\ref{fig6}b.

After the therapy activation, the cancerous cells initially respond
to the medicine, starting a decreasing trend towards a total
disappearance. When the mutation rates start to grow, the number of
cancerous cells starts to increase again, exceeding the previous
therapy activation level and quickly reaching the total value N.
Fig.~\ref{fig6}b represents a clear example of acquired resistance
that brings the system to a totally cancerous cell condition in
spite of the presence of therapy.

\section{Conclusions}\label{concl}
Chronic myeloid leukemia is a blood cancer that causes an
overproduction of immature white cells. In this paper we provide a
statistical description of the effects of an imatinib-like based
therapy on the evolutionary dynamics of normal, first-mutant and
cancerous cell populations in a stochastic model of chronic myeloid
leukemia. Imatinib strongly limits the reproduction of only
cancerous cells, permitting a cancer targeted therapy.
Unfortunately, acquired drug resistance is the major limitation for
the successful treatment of cancer. In fact, several patients,
developing resistance to the therapy, show a relapse towards the
accelerated phase and blast crisis which are strongly characterized
by genetic mutations.

In this work, we study the effect of different mutation rates on the
somatic evolution of cancer. We find that the response to the
therapy, in terms of a permanent reduction of malignant cells,
depends on both the inhibitory capability of the medicine and on the
levels of mutation rates from normal cells to leukemic types. In the
best cases, a total (but temporary) recovery is usually obtained in
almost 100 days, as experimentally observed in clinical cases.
Nevertheless, we show that an enhancement of the mutation rates from
healthy cells to type 1 (first-mutants) and from type 1 to cancerous
cells, caused by genetic alterations, can retard the complete
depletion of leukemic cells. Our modeled evolutionary dynamics of
leukemic cells are in agreement with the different temporal response
of patients treated with imatinib observed in clinical studies, as
summarized in Fig.~1b of Ref.~\cite{Roe2006}.

Moreover, we point out that, in the worst cases, this enhancement of
mutation rates can bring the cell system to an acquired resistance
to the therapy. In fact, after an initial response, we find a
dramatic enhancement of leukemic cell abundance even in the presence
of a significant reduction of cancerous cell fitness induced by the
therapy. Our theoretical results are further supported by the
experimental evidence of an increase of genetic mutations in
patients which show acquired resistance \cite{Sov2006}.

Our work provides a step towards the construction of a comprehensive
theory for the dynamics of cancer evolution. A more realistic
description of therapeutic scenarios, however, needs to take into
account (a) the therapy effect on the leukemic cell death rate, (b)
tunneling phenomena from type 0 to type 2 cells and (c) a dose
dependent response of patients to the therapy. All these topics will
be investigated in a next work.

\section*{Acknowledgments}
This work was supported by MIUR and CNISM-INFM.

\end{document}